\documentstyle[12pt,epsfig]{article}
\begin{document}
\begin{center}
\vskip 2.0 truecm
{\bf \Large  Modified  negative binomial description of the \\
            multiplicity distributions in lepton-nucleon
            scattering\footnote{This 
            work was partially supported 
            by INTAS, contract INTAS-93-3602 } }
\vskip 1.5 truecm
 {\sc   O. G. Tchikilev\footnote{ E-mail: tchikilov@mx.ihep.su} }
\vskip 0.4cm
  {\em Institute for High Energy Physics  \\
   142284, Protvino, Russia\/}
\end{center}
\vskip 1.5 truecm
\begin{abstract}
\noindent
It is shown that  charged hadron multiplicity distributions
in lepton-nucleon scattering are fairly well described by the
modified negative binomial distribution in the energy range from
   $3-4$ to $\simeq 220$~GeV. The energy behaviour of the parameter 
 $k$ is similar to the dependence observed for $e^{+}e^{-}$ annihilation.
\end{abstract}
\vskip 1.5cm
\begin{center}
(  to be published in  Phys. Lett. B)
\end{center}

\newpage
\pagestyle{plain}
\subsection*{ }
 In recent papers~[1,2,3,4,5] 
negatively charged particle multiplicity
 distributions in $e^+e^-$ annihilation have been well described by the 
 modified negative binomial distribution (MNBD) with the parameter
 $ \Delta $ practically energy independent and the parameter 
 $k$ approaching the value 7.
 The aim of this letter is to show that the MNBD fairly well
 describes multiplicity 
 distributions in lepton-nucleon interactions with similar behaviour of the
 parameters.

 Let us remind that the
 MNBD can be defined by the probability generating function
\begin{equation} 
           M(x) = \sum_{n}{P_{n}x^{n}} = 
   {\biggl({{1+\Delta\, (1-x)}\over{1+r\,(1-x)}}\biggr)}^{k},
\end{equation}
where  $P_n$ is the probability to produce ~$n$~ particles. 
For integer $k$, positive $r$ and negative $\Delta$ lying in the interval
$-1 \leq \Delta \leq 0$ the MNBD is
positive definite and can be computed as a convolution of the Newton
 binomial (the numerator in (1)) and the negative binomial (the denominator
in (1)),  these two  distributions are also positive
definite. The probabilities
 $P_{n}$ for the MNBD can be computed also using iteration relations given
 in the paper~[4] or the formulae given in the paper~[6].

 In tables 1 and 2 the results of the MNBD fits 
with fixed $\Delta=-0.5$
to the published negatively
 charged particle multiplicity distributions for the reactions
 $\mu^{\pm}+p\rightarrow\mu^{\pm}+X^+$~[7,8,9] and 
 $\mu^{+}+D_2\rightarrow\mu^{+}+X^+$~[10]
 are given for different
 intervals of the invariant mass $W$  of the hadronic system $X$ and the
 momentum transfer squared $Q^2$. 
 The parameter $r$ has been replaced in fits by the
  the mean multiplicity
 $<n>$ using the constraint
%
\begin{equation}
             r = \Delta + <n>/k.
\end{equation}
\noindent
 The data have been selected using two simple criteria: a) the number of
 measured points $n_{p}$ in the multiplicity distribution exceeds 3 and b) 
  $W~>~3~GeV$.  For $\mu^+D_2$ reaction  the published uncorrected
 multiplicity distributions have been used, since 
 the effect of corrections is rather small as can be seen from the comparison
 of the uncorrected multiplicity distributions with results of the
 negative binomial fits to the ``unfolded'' distributions~[10,11].

   One can see from tables 1 and 2 that the quality of fits is quite qood,
 the points with big $\chi^2/NDF$ values are mainly at small $W$ and/or $Q^2$.
 Big $\chi^2/NDF$ at low $W$ can be explained by the rise of the
 parameter $k$ from $\simeq 1$ to $\simeq 3$ in the wide $W$ intervals,
 at higher $W$ the relative variation of the parameter $k$ becomes smaller
 and therefore its influence is not seen. 
  It is necessary to note that for points with big $\chi^2/NDF$ 
 acceptable fits are obtained with free $\Delta$ (not shown). The reasonable
 fits with free $\Delta$ are obtained also in the energy range
  $W<3$~GeV, where
 $n_{p}$ is usually below 4, i.e. the number of degrees of freedom is too
 small for two-parameter fits.

  In the table~3 the results of the MNBD fits to the multiplicity distributions
 in the reactions $\nu + p \rightarrow \mu^- + X^{++}$~[12,13],
 $\nu + n \rightarrow \mu^- + X^+$~[12] and
 $\overline{\nu} + p \rightarrow \mu^+ + X^0$~[14] are given. For these
 reactions the quality of fits is also quite qood, big $\chi^2/NDF$ again
 are observed mainly for the $W$ intervals near $3-4$~GeV.
  
 As seen from tables 1 and 2 the $Q^2$ dependence of the parameter $k$ 
 for $\mu p$ reactions in
 different $W$ intervals is weak or even absent. 
 The energy dependence of the parameter $k$ for the studied lepton-hadron 
 processes is compared in fig.~1 with the energy dependence of the parameter
 $k$ for $e^+e^-$ annihilation, the $k$ values for $e^+e^-$ data were
 taken from the papers~[4,5]. 
 For clarity only the part of the values given
 in tables 1-3 is shown in this figure. The errors of the parameter $k$
 for the studied reactions (not given) are of the order $\simeq ~1$ and
 exceed the errors for statistically more significant $e^+e^-$ data.
 The $k$ values for $\mu p$ reactions
 follow the dependence observed for $e^+e^-$ annihilation. The $k$ values for
 $\nu p$ scattering are slightly below the ones for $e^+e^-$ annihilation and
 approach them from below with the energy increase whereas the $k$ values
 for $\overline{\nu}p$ scattering are slightly above the $e^+e^-$ data.
  The $k$ values for $\mu^+ D_2$ scattering are
 slightly below the $k$ values for $\mu^+ p$ scattering.

 Recently  charged particle multiplicity distributions for the reaction 
 $e^+ + p \rightarrow e^+ + X$ have been published by the H1 
 Collaboration~[15]. The results of the MNBD fits with the
 fixed $k=7$, as suggested by the $e^+e^-$ data, are given in the
 table~4 for the maximum available pseudorapidity interval
 $1< \eta^* <5$. The $\chi^2/NDF$ values for MNBD fits are good, but they
 should be considered just indicative, since the full covariance matrix
 is not published and therefore the proper treatment of the correlations
 between measurements of the neighbour multiplicities is not possible.
 This contrasts  presented earlier fits of the muon and (anti)neutrino data
 obtained mainly using bubble and streamer chambers, where these
 correlations are small. Nevertheless one can compare these values with
 the $\chi^2/NDF$, obtained for fits using another generalization of the
 negative binomial,  so called Generalized Negative Binomial 
 Distribution~(GNBD)~[16,17] (see table~4). 
 The $\chi^2/NDF$ values for MNBD fits are
 significantly smaller than for GNBD fits.
  It is necessary also to note that MNBD fits are 
 one-parameter fits, since only $\Delta$ is free parameter and $r$ has been
 fixed using the measured mean multiplicity $<n_{ch}>$, whereas the
 GNBD fits are two-parameter fits.

  One can also estimate the parameter $k$ from the ratio of the dispersion 
 $D = (<n^2>-<n>^2)^{1/2}$ to the mean multiplicity $<n>$ using formula
\begin{equation}
   \frac{D^2}{<n>^2} = \frac{1}{k} + \frac{1+2\Delta}{<n>},
\end{equation}
 for $\Delta = -0.5$ it gives
\begin{equation}
   k = \frac{<n>^2}{D^2}.
\end{equation}

 This formula has been applied to the data given in the papers~[18,19,20], 
 where multiplicity distributions are not tabulated.
 The ratio of the
 dispersion $D_{ch}$ to the 
 mean multiplicity for charged particle multiplicity distribution 
  has been
 calculated from the results of the negative binomial fits
 (parameters $k_{NB}$ and $<n>_{NB}$) and has been
 transformed to the ratio for negative charged particle multiplicity
 distribution. The final formula for the parameter $k$~(called $\kappa$ to be
 distinguished from $k$ for MNBD fits) is
\begin{equation}
  \kappa = (\frac{k_{NB} <n>_{NB}}{k_{NB}+<n>_{NB}})
  (\frac{<n>_{NB} - |q|}{<n>_{NB}})^2,
\end{equation}
 where q is the total charge of the hadronic system $X$ in the studied 
 reactions.

 The energy dependence of the parameter $\kappa$  shown in the fig.~2 is
similar (if not coincides) with the energy dependence of the parameter
 $k$ in the fig.~1. The $\kappa$ values for the H1 data, shown in the
figure were calculated
 from the values of
 the second normalized moment $C_{2} = <n^2>/<n>^2$ given in
 \cite{h1} for the interval $1<\eta^*<5$ 
  using the relation $(<n>/D)^2 = 1/(C_2-1)$.
 The $C_2$ for the charged particle multiplicity distribution are used,
since at the HERA energies the effect of the total charge $q$ 
 of the system $X$ (second factor
 in the formula 5) can be neglected.

 What is the physical meaning of the MNBD parameters $k$ and $\Delta$?
 One can assume that $k$ is the number of sources of particle production
 at some stage of interaction, each source develops according to some
 branching process producing intermediate neutral clusters
 (as proposed in \cite{ch1,ch2,ch3})\footnote{It is of interest to note that
 the parameter $k$ from MNBD fits is rather close to the parameter $c$
 of the Levy parametrization for the KNO scaling function 
 $\psi (z)$\cite{na5,sold,jones}. The possible relation between Levy and
 MNBD parametrizations has been mentioned in~\cite{tch}.}.
The constancy of the parameter
 $\Delta$ indicates that this branching process is pure birth branching
 process and in this case $-\Delta$ represents the cluster decay
 probability into charged hadron pair ($1+\Delta$ is the probabilty of
 cluster decay into pair of neutral hadrons). Following the 
 statistical model proposed by Goulianos \cite{goul} 
 it was assumed in \cite{ch1,ch2,ch3} that $\Delta$
 should be equal -0.5. But this assumption
 is incorrect, since it predicts wrong ratio
 of the average
 neutral and charged pion multiplicities. In the Goulianos model this
 deficiency was corrected
 by ``asymmetric'' assumption that the neutral cluster
 decays into charged pion pair or
 transforms
 into one neutral hadron. The observed neutral/charged ratio is restored 
 in the model with $2\pi^0$ cluster decay mode if~
 $\Delta=-2/3$, this value is 
 close to the value $\Delta= -0.76$ used in the MNBD fits of the
 $e^+e^-$ data. One should note that the
  value $\Delta=-2/3$ is expected if neutral clusters have zero
 isotopic spin. The fits of the lepton-nucleon data with $\Delta$
 equal to -2/3 or -0.76 give also acceptable fits but in general worse
 compared to the $\Delta=-0.5$.
 
 In conclusion, it is shown that the MNBD with fixed parameter $\Delta=-0.5$
 quite well describes lepton-nucleon multiplicity data in the wide energy
 range from $3-4$ to $\simeq$ 220~GeV. The parameter $k$ for lepton-nucleon
 interactions has the energy dependence similar to the energy dependence for
 $e^+e^-$ annihilation. The evidence, given in this paper,
 for the same asymptotic value $k=7$ for $e^+e^-$ annihilation and 
 lepton-nucleon scattering has no direct theoretical explanation now.
 Future experimental and theoretical studies should show whether this
 coincidence
 is accidental or reflects some common dynamical mechanism.

\subsection*{Acknowledgements}
\vskip 3mm

 I thank S. S\"oldner-Rembold and W. Wittek for detailed information
 on multiplicity distributions in the experiment E665.

\newpage

\newpage
\begin{center}
\Large{\underline{Figure Captions}}
\end{center}
\vskip 2truecm


 Fig.1 The energy dependence of the parameter $k$ obtained from the MNBD fits
 with fixed $\Delta=-0.5$
  to the negative charged particle multiplicity distributions in the
 lepton-nucleon interactions compared to the $k$ values for $e^{+}e^{-}$
  annihilation. For clarity some points are slightly shifted from
 integer values. The histograms are drawn to guide the eye.
\vskip 0.5truecm
 Fig.2 The energy dependence of the parameter 
$\kappa = (<n>/D)^{2} =1/(C_{2}-1)$ for the reactions $e^+ +p \rightarrow
e^+ + X$~\cite{h1}, $\mu^+ +p \rightarrow \mu^+ +X$~\cite{arneo},
 $\overline{\nu} + p \rightarrow \mu^+ +X$~\cite{jong,jones} and
 $\nu + p \rightarrow \mu^- +X$~\cite{jong,jones} compared to the $k$
 values for $e^+e^-$ annihilation. For the H1 data the measurements in the
 pseudorapidity interval $1 < \eta^* < 5$ are used.

\newpage
\newcommand{\qq}{\mbox{$Q^{2}$}}
\newcommand{\en}{\mbox{$\sqrt s$}}
\newcommand{\np}{\mbox{$n_p$}}
\newcommand{\pnch}{\mbox{$P(n_{ch})$}}
\newcommand{\nch}{\mbox{$n_{ch}$}}
\newcommand{\avm}{\mbox{$<n_{-}>$}}
\newcommand{\avn}{\mbox{$<n_{-}>$(fit)}}
\newcommand{\axi}{\mbox{$\chi^2$/NDF}}
\newcommand{\dda}{\mbox{$\times 10^{-1}$}}
\newcommand{\ddb}{\mbox{$\times 10^{-2}$}}
\newcommand{\ddc}{\mbox{$\times 10^{-3}$}}
\newcommand{\ddd}{\mbox{$\times 10^{-4}$}}
\newcommand{\dde}{\mbox{$\times 10^{-5}$}}
\newcommand{\ddf}{\mbox{$\times 10^{-6}$}}

\begin{table}[bth]
\caption{Results of the modified negative binomial fits with fixed
 parameter 
         \mbox{$\Delta=-0.5$} to the 
         negative charged particle multiplicity distributions 
         in  lepton-hadron reactions 
\mbox{$\mu^+ +p \rightarrow \mu^+ +X$}~[9],  
           \mbox{$\mu^- +p \rightarrow \mu^- +X$}~[7] and
 \mbox{$\mu^+ +D_2 \rightarrow \mu^+ + X$}~[10].}
                  
\begin{center}
\begin{tabular}{|c|c|c|c|c|c|c|}
\hline\hline             
Experiment       & W $(GeV)$  & \qq $(GeV)^2$  & \avn  &  k     & \axi \\
\hline\hline
  &  & 0.24 &1.146$\pm 0.018$ & 2
                                                   &10.9/3\\
   &2.8$\div$3.8  & 0.54 &1.017$\pm 0.018$ &2   &7.0/3\\
 &  & 0.95 & 1.071$\pm 0.024$ &2  &7.0/3 \\
J.Ballam~\cite{bal} & &1.80  &0.782$\pm 0.076$  &1 &13.1/3 \\ \cline{2-6}
$\mu^-p$  &   & 0.0675  &1.222$\pm 0.029$  & 3 &30.4/3 \\
16~GeV/c  & 3.8$\div$5.0 & 0.25  &1.181$\pm 0.075$ &2 & 5.8/3 \\
  &  & 0.56  &1.354$\pm 0.131$ &2 & 0.8/3 \\
  &  & 1.38  &1.151$\pm 0.056$ &3 & 0.2/3 \\
\hline
  &4$\div$6  &   &1.754$\pm 0.071$ & 4    &1.3/5 \\
    & 6$\div$8  &  &2.067$\pm 0.086$  & 3   &3.3/5 \\
 NA9 \cite{na5}       & 8$\div$10  & &2.339$\pm 0.037$  & 4
                                                   &9.4/6 \\
 $\mu^+p$ & 10$\div$12  & 4$\div$200 &2.598$\pm 0.080$ & 5    &19.2/6 \\
 280~GeV/c            & 12$\div$14  & &2.848$\pm 0.083$ & 5   &6.8/6\\
                 & 14$\div$16  &  & 3.011$\pm 0.072$ & 6
                                                   &7.2/7 \\
                 & 16$\div$18  &  & 3.108$\pm 0.081$ & 6   &19.1/8 \\
                 & 18$\div$20  &  & 3.206$\pm 0.068$ & 7
                                                   &39.6/8\\
\hline
  & 8$\div$10 &  & 4.984$\pm 0.108$ & 4 & 4.9/14 \\
  & 10$\div$-12 & & 5.474$\pm 0.095$ & 4 & 17.6/15 \\
  & 12$\div$-14 & & 5.943 $\pm 0.106$ &  4 & 33.3/17 \\
 E665~\cite{e665} & 14$\div$16 & & 6.525 $\pm 0.115$ & 4 & 21.4/19 \\
 $\mu^+D_2$ & 16$\div$18 &$>1.0$ & 6.787 $\pm 0.112$ & 4 & 36.4/21 \\
 490~GeV/c & 18$\div$20 & & 7.336 $\pm 0.131$ & 4 & 30.4/19 \\
           & 20$\div$22 & & 7.241 $\pm 0.131$ & 5 & 22.2/21 \\
           & 22$\div$25 & & 7.813 $\pm 0.135$ & 4 & 36.4/23 \\
           & 25$\div$30 & & 7.861 $\pm 0.156$ & 4 & 28.3/22 \\
\hline            
\end{tabular}
\end{center}
\end{table}
\newpage
\vskip 2truecm
\begin{table}[bth]
\caption{The same as in Table~1 for 
 reaction \mbox{$\mu^+ +p \rightarrow \mu^+ +X$}  at
 14~GeV/c~[8].}
                  
\begin{center}
\begin{tabular}{|c|c|c|c|c|c|}
\hline\hline 
 W $(GeV)$  &\qq $(GeV)^2$  & \avn  &  k     & \axi \\
\hline\hline
 3.1  &   &1.072$\pm 0.075$ & 2    &8.0/3 \\
 3.5  &  &1.328$\pm 0.135$  & 3   &3.0/3 \\
 3.7  & 0.3$\div$0.5 &1.225$\pm 0.117$  & 2
                                                   &2.7/3 \\
 3.9  &  &1.325$\pm 0.138$ & 2    &14.2/3 \\
 4.3  & &1.378$\pm 0.116$ & 4   &6.8/3\\
\hline
 3.1  &  & 1.002$\pm 0.036$ & 2
                                                   &13.8/3 \\
 3.3  &  & 1.094$\pm 0.070$ & 2   &0.1/3 \\
 3.5  & 0.68 & 1.028$\pm 0.069$ & 2
                                                   &7.2/3\\
 3.7   &  &1.207$\pm 0.055$ & 3
                                                   &0.9/3\\
 3.9  &  &1.203$\pm 0.079$ &2   &2.6/3\\
 4.3  & & 1.368$\pm 0.131$ &2  &8.6/3 \\
\hline
 3.1 &  &1.405$\pm 0.052$  &2 &12.4/3 \\
 3.3  &   &1.119$\pm 0.024$  & 3 &1.3/3 \\
 3.5 & 1.75   &1.061$\pm 0.038$ &2 & 10.0/3 \\
 3.7 &    &1.150$\pm 0.075$ &2 & 6.3/3 \\
 3.9 &    &1.141$\pm 0.180$ &1 & 1.5/3 \\
 4.3 &    &1.328$\pm 0.121$ &2 & 3.7/3 \\
\hline
          &0.3$\div$0.5&1.042 $\pm 0.037$ & 2 & 25.3/3 \\
          &0.5$\div$1.0&1.002 $\pm 0.033$ & 2 & 17.4/3 \\
 2.8$\div$3.6 &1.0$\div$1.5&0.987 $\pm 0.033$ & 2 &  7.7/3 \\
          &1.5$\div$2.0&1.027 $\pm 0.037$ & 2 & 11.3/3 \\
          &2.0$\div$4.5&1.054 $\pm 0.038$ & 2 &  1.4/3 \\
\hline
          &0.3$\div$0.5&1.296 $\pm 0.074$ & 3 & 23.7/3 \\
          &0.5$\div$1.0&1.358 $\pm 0.074$ & 2 & 10.2/3 \\
 3.6$\div$4.7  &1.0$\div$1.5&1.163 $\pm 0.052$ & 2 &  4.5/3 \\
          &1.5$\div$2.0&1.279 $\pm 0.105$ & 3 &  1.4/3 \\
          &2.0$\div$4.5&1.261 $\pm 0.147$ & 2 &  4.8/3 \\ 
\hline            
\end{tabular}
\end{center}
\end{table}
\newpage
\vskip 2truecm
\begin{table}[bth]
\caption{The same as in Table~1 for \mbox{$\nu p \rightarrow \mu^-
 +X^{++}$}~[12,13], 
\mbox{$\nu n \rightarrow \mu^{-} +X^+$}~[12]     
  and \mbox{$\overline{\nu}p \rightarrow \mu^+ +X^0$}~[14] reactions.} 
\begin{center}
\begin{tabular}{|c|c|c|c|c|c|}
\hline\hline 
Experiment       & W $(GeV)$    & \avn  &  k     & \axi \\
\hline\hline
         & 2.83$\div$3.46    &0.766$\pm 0.022$  & 1   &32.6/3 \\
        & 3.46$\div$4.0  &1.130$\pm 0.024$  & 2
                                                   &3.0/5 \\
                 & 4.0$\div$4.8    &1.287$\pm 0.030$ & 2    &21.8/5 \\
 $\nu p$ & 4.8$\div$5.66   &1.606$\pm 0.036$ & 3   &5.7/5\\
D.~Zieminska~\cite{ziem}  & 5.66$\div$6.76    & 1.860$\pm 0.047$ & 3
                                                   &12.4/6 \\
                 & 6.76$\div$7.94    & 2.048$\pm 0.068$ & 3   &13.2/5 \\
                 & 7.94$\div$9.49    & 2.356$\pm 0.078$ & 4
                                                   &9.7/6\\
                 & 9.49$\div$11.18   & 2.642$\pm 0.124$ & 4
                                                   &13.1/6\\
  & 11.18$\div$15.0  &2.710$\pm 0.151$ &4   &3.8/6\\
\hline
  & 2.83$\div$3.46 & 1.036$\pm 0.012$ &2  &24.0/3 \\
  & 3.46$\div$4.0  & 1.185$\pm 0.020$ &2  &25.2/4 \\
 &4.0$\div$4.8   &1.560$\pm 0.020$  &3 &13.0/5 \\
 $\nu n$ & 4.8$\div$5.66    &1.721$\pm 0.030$  & 3 &8.1/5 \\
D.~Zieminska~\cite{ziem} & 5.66$\div$6.76 &1.990$\pm 0.048$ &3 & 3.7/5 \\
  & 6.76$\div$7.94  &2.217$\pm 0.054$ &4 & 1.7/5 \\
  & 7.94$\div$9.49  &2.476$\pm 0.076$ &4 & 3.4/6 \\
  & 9.49$\div$11.18 &2.004$\pm 0.036$ &5 & 9.8/6 \\
  & 11.18$\div$15.0 &3.128$\pm 0.187$ &3 & 5.8/6 \\
\hline
 & 3.16$\div$4.47 & 1.065 $\pm 0.065$ & 2 & 2.8/3 \\
$\nu p$ & 4.47$\div$5.48 & 1.279 $\pm 0.085$ & 2 & 2.8/4 \\
J.~Chapman~\cite{chap}& 5.48$\div$7.07 & 1.776 $\pm 0.102$ & 3 & 15.3/4 \\
    & 7.07$\div$10   & 1.564 $\pm 0.130$ & 2 & 9.0/6 \\
\hline
 & 3.0$\div$3.5 & 1.196 $\pm 0.042$ & 2 & 7.5/3 \\
    & 3.5$\div$4.0 & 1.545 $\pm 0.033$ & 3 & 9.4/3 \\
$\overline{\nu} p$& 4.0$\div$4.5 & 1.864 $\pm 0.069$ & 4 & 3.7/3 \\
M.~Derrick~\cite{der}  & 4.5$\div$5.0 & 1.899 $\pm 0.088$ & 4 & 3.4/3 \\
        & 5.0$\div$7.5 & 2.510 $\pm 0.073$ & 5 & 13.3/4 \\
        & 7.5$\div$10.0 & 2.840 $\pm 0.175$ & 8 & 0.3/4 \\
\hline            
\end{tabular}
\end{center}
\end{table}
\vskip 2truecm
\begin{table}[bth]
\caption{The results of the MNBD fits with $k=7$ to the
 H1 data~[15]  with 
 $<n_{ch}>$ fixed at the published values for the pseudorapidity
interval $1 < \eta^* < 5$ and for the $Q^2$ interval
$10 < Q^2 < 1000$~(Gev/c)$^2$. In last column the $\chi^2/NDF$ ~for
 fits using
 Generalized Negative Binomial Distribution from the paper~[17]
 are shown.}
                  
\begin{center}
\begin{tabular}{|c|c|c|c|}
\hline\hline
 W (Gev)   & -$\Delta$  & \axi  & \axi (from~[17])    \\ 
\hline\hline
80$\div$115  &0.216 $\pm 0.034$ & 2.7/18 & 12/17    \\
115$\div$150  &0.233 $\pm 0.036$ & 10.6/21 & 26/20    \\
150$\div$185  &0.187 $\pm 0.049$ & 4.6/22 & 19/21      \\
185$\div$220  &0.264 $\pm 0.057$ & 11.3/23 & 16/22      \\
\hline            
\end{tabular}
\end{center}
\end{table}
\newpage
\begin{figure}[th]
\begin{center}
\epsfig{file=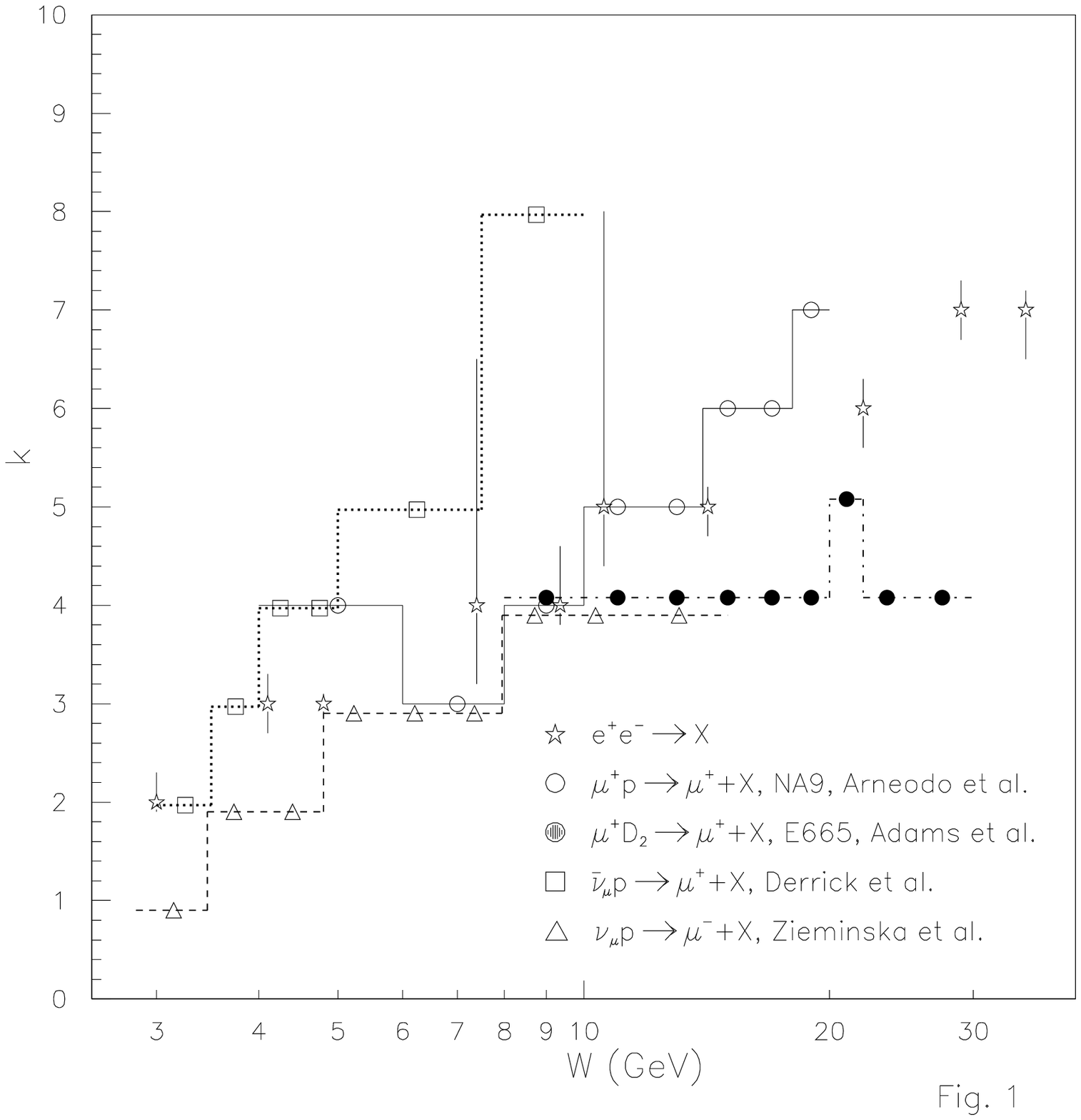,bbllx=85pt,bblly=-5pt,bburx=590pt,bbury=625pt,%
width=18cm,clip=}
\end{center}
\end{figure}
\newpage
\begin{figure}[th]
\begin{center}
\epsfig{file=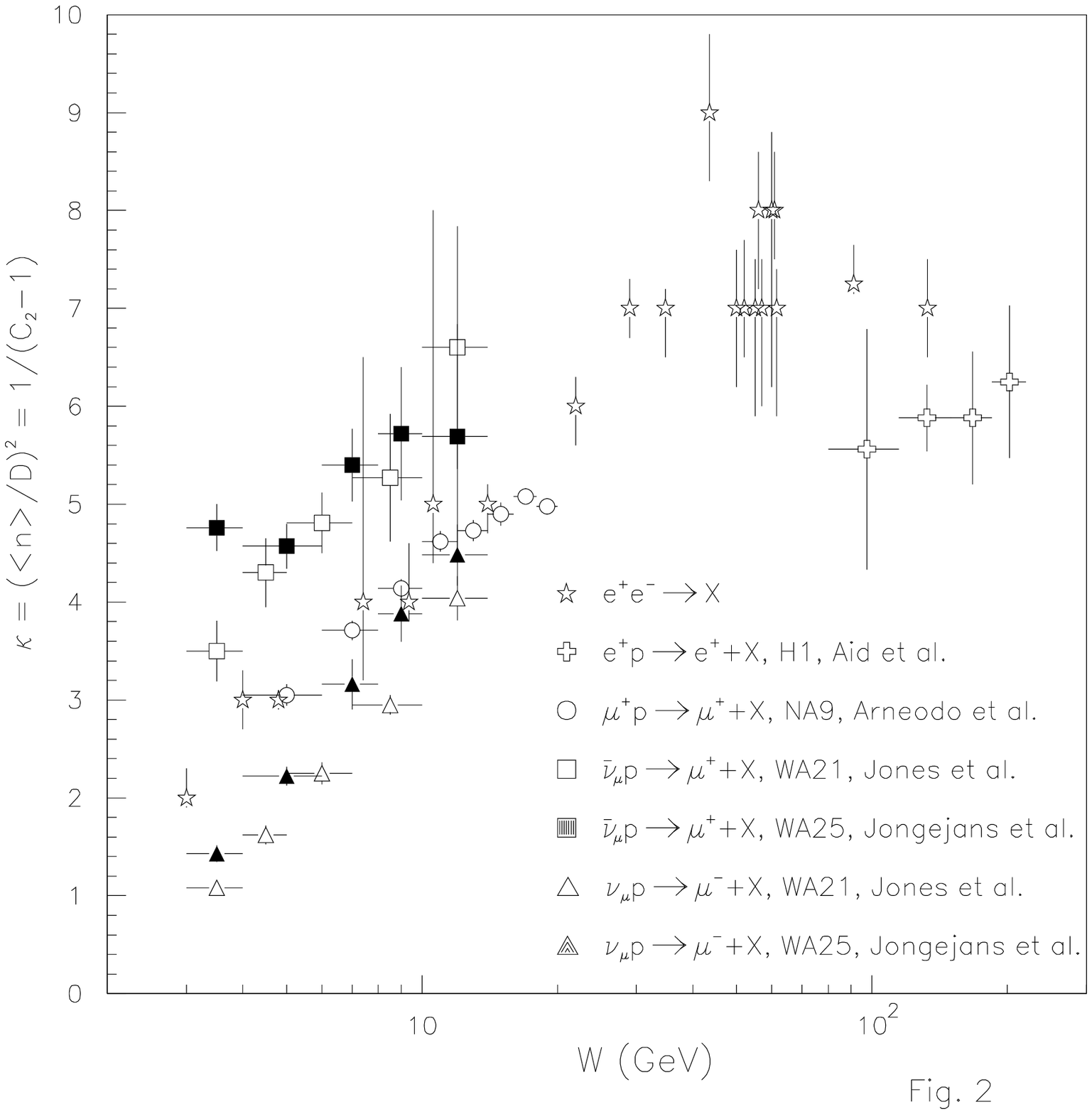,bbllx=85pt,bblly=-5pt,bburx=590pt,bbury=625pt,%
width=18cm,clip=}
\end{center}
\end{figure}

\end{document}